\documentclass[10pt, letterpaper]{article}

\RequirePackage[left=1in,right=1in,top=1in,bottom=1in]{geometry}

\usepackage{modular}
\usepackage{graphicx}
\usepackage{epstopdf}
\usepackage{multicol}
\usepackage{multirow}

\usepackage{color}
\usepackage{siunitx}
\usepackage{multirow}
\usepackage{algorithm}  
\usepackage{algorithmic}  
\usepackage{amsmath}
\usepackage{makecell}
\usepackage{mathrsfs}
\usepackage{amsfonts}
\usepackage{hyphenat}
\usepackage{scrextend}
\usepackage{hyperref}
\usepackage{authblk}
\usepackage{cite}
\usepackage{amsthm}
\usepackage[utf8]{inputenc}
\usepackage{soul}

\title{Polar Codes-based Information Reconciliation Scheme with Frozen Bits Erasure Strategy for Quantum Key Distribution}

\author[1]{Bang-Ying Tang}
\author[2]{Chun-Qing Wu}
\author[1]{Wei Peng}
\author[3,\thanks{Corresponding authors: liubo08@nudt.edu.cn, wlyu@nudt.edu.cn}]{Bo Liu}
\author[1,$^ *$]{Wan-Rong Yu}

{
\affil[1]{\small College of Computer Science and Technology, National University of Defense Technology, Changsha 410073, China}
\affil[2]{School of Electronics and Communication Engineering, Sun Yat-sen University, Shenzhen 518100, China}
\affil[3]{College of Advanced Interdisciplinary Studies, National University of Defense Technology, Changsha 410073, China}
}
\date{}

\begin{document}
\maketitle
\begin{abstract}
Information reconciliation (IR) ensures the correctness of quantum key distribution systems, by correcting the error bits existed in the sifted keys. In this article, we propose a polar codes-based IR scheme with the frozen bits erasure strategy, where an equivalent transmission of sifted keys is conducted, so that the frozen bits in the decoding procedure is erased to 0. Thus, our IR scheme can be implemented efficiently without the assumption of true random numbers. Furthermore, we implement the proposed IR scheme with the fast simplified successive cancellation list decoder and its throughput reaches to \SI{0.88}{Mbps} with the yield of $0.8333$, where the decoder list size is $16$, the block size is \SI{1}{Mb} and the quantum bit error rate is $0.02$.
\end{abstract}

\begin{multicols}{2} 

\section{Introduction}

Quantum  key distribution (QKD) can provide the information-theoretical-secure keys for distant users~\cite{RN4}. In realistic QKD systems, the sifted keys ($K_{s}^A$ and $K_{s}^B$) with error bits are generated for users (Alice and Bob) after the quantum communication phase and basis sifting procedure~\cite{RN624}. Information reconciliation (IR) ensures the correctness of QKD systems, by correcting these error bits with exchanged syndrome information via the classical channel and finally gains the symmetric bit strings $K_\textrm{IR}$. Furthermore, IR is widely applied to various secure communication scenarios, such as physical layer security~\cite{RN493,RN496,RN497}, underwater acoustic communication~\cite{RN407,RN406}, and so on.

IR schemes are mainly performed by the interactive primitives~\cite{RN148, RN115} or the forward error correction (FEC) codes, \textit{e.g.} Low-density parity-check (LDPC) codes, polar codes~\cite{ RN159, RN156, RN129}. Although reaches high efficiency, interactive primitives-based IR schemes (BBBSS and Cascade) have limited applications due to the heavy communication latency~\cite{RN159}. LDPC codes-based IR scheme is widely applied in the QKD systems, but its efficiency and applications are limited by the check matrix corresponding to each quantum bit error rate (QBER)~\cite{RN95, RN159}. Recently, a polar codes-based IR scheme has been proposed to further improve the efficiency~\cite{RN122, RN125}.

The first polar codes-based IR scheme was proposed by Jouguet and Kunz-Jacques and reached the efficiency of $1.121$ and failure probability of $0.08$, with successive cancellation (SC) decoder when the QBER is $0.02$ and the block size is \SI{16}{Mb}~\cite{RN129}. Then, three configuration strategies were developed to adapt the polar codes into IR schemes: direct decoding (DD) strategy, bit flipping decoding (BFD) strategy, and length-adaptive BFD strategy~\cite{RN98}. The DD strategy and the BFD strategy, have the higher efficiency than the length-adaptive BFD strategy, are widely used in the polar codes-based IR schemes. With the DD strategy, the polar codes-based IR scheme was further performed in QKD systems and achieved the efficiency of $1.176$ and the failure probability of $0.001$ with the successive cancellation list (SCL) decoder, the list size of $16$, and the QBER of $0.02$~\cite{RN99, RN98}. In our previous work, we proposed the polar codes-based feedback IR scheme (with the BFD strategy), which decreased the failure probability to $10^{-8}$ with the efficiency of $1.055$ when the list size of SCL decoder is $16$, the block size is \SI{1}{Gb} and the QBER is $0.02$~\cite{RN676}. Nevertheless, the throughput of the polar codes-based IR scheme is limited by inefficient implementations. The polar codes-based IR scheme with the BFD strategy could be implemented with the efficient decoder whose frozen bits are constant (usually fixed to 0), such as simplified SC decoder, fast simplified SCL decoder and so on~\cite{RN98, RN111}. However, true random numbers (TRNs) are indispensable to the polar codes-based IR scheme with BFD strategy, would increase the complexity of the practical systems and might open security loopholes with the inappropriate implementation. 

In this article, we propose the polar codes-based IR scheme with the frozen bits erasure (FBE) strategy, which can be implemented efficiently without the TRNs. The proposed IR scheme mainly contains two phases: the equivalent transmission of sifted keys with FBE strategy and the error bits correction of equivalent sifted keys. In the former, Alice distills the syndrome vector $W$ and sends $W$ to Bob via the classical channel. Alice and Bob both conduct the ``XOR'' operation between the sifted keys and the encoded vector of $W$, so as to generate a new couple of vectors ($X$ and $X'$). In the latter, Alice extracts the key $K^A_\textrm{IR}$ from the encoded $X$ and sends the cyclic redundancy check (CRC) value of $K^A_\textrm{IR}$ to Bob. Bob decodes the generated vector $X'$ with the frozen bits of $0$ and the received CRC value to $U'$. Finally, Bob extracts the key $K^B_\textrm{IR}$ from $U'$. The polar codes-based IR scheme with the FBE strategy could be efficiently implemented on the commercial computer without the extra hardware. We implemented the IR scheme with the fast simplified SCL (FSSCL) decoder on the commercial computer. The implementation reaches the throughput of \SI{0.88}{Mbps} and the yield of $0.8333$ (efficiency of $1.760$ and failure probability of $0.0004$) with the decoder list size of $16$, the block size of \SI{1}{Mb} and QBER of $0.02$. Thus, the IR scheme could be applied in the practical QKD systems, especially with the ultra worse link conditions.

\section{Information Reconciliation}
In quantum key distribution (QKD) systems, the communication parties (Alice and Bob) gain the sifted keys ($K^{A}_{s}$ and $K^{B}_{s}$) of length $n$ with the quantum bits error rate (QBER) $E_\mu$ after the quantum communication phase and basis/key sifting procedures. 

Then, Alice and Bob exchange the syndrome string $S$ via the classical channel and correct the sifted keys to the weak secure keys ($K_\mathrm{IR} ^ A$ and $K_\mathrm{IR} ^ B$) in the information reconciliation (IR) procedure. The failure probability $\varepsilon$ of IR represents the correctness of IR as
\begin{equation}
	\varepsilon \geq \operatorname{Pr}(K_\mathrm{IR} ^ A \neq K_\mathrm{IR} ^ B).
\end{equation}

Assume $K_\mathrm{IR} = K_\mathrm{IR} ^ A ( K_\mathrm{IR} ^ B)$, when IR procedure is conducted successfully. The syndrome $S$ through the classical channel discloses partial information of the key and decreases the secure key rate of systems. The leaked information is represented by the efficiency of IR, which is difined as 
\begin{equation}
	f(E_\mu)=\frac{1-H_2(K_\mathrm{IR}|S)}{H_2(E_\mu)},
\end{equation}
where $H_2(x)$ is the binary Shannon entropy as 
\begin{equation}
	H_2(x)=-x \log _2 (x) - (1-x) \log _2 (1-x).
\end{equation}

Finally, the yield $\gamma$ of each sifted bit evaluates the performance of IR scheme and is calculated as
\begin{equation}
	\label{equ:gamma}
	\gamma = \left( 1-\varepsilon \right) \left[ 1-f \left( E_\mu \right) H_2 \left( E_\mu \right) \right].
\end{equation}

\section{Polar codes-based IR scheme with FBE Strategy}

Polar codes are adapted into information reconciliation (IR) for the advantages: the potential to achieve Shannon-limit efficiency, low complexity $O(n \log n)$ of encoding and decoding procedure~\cite{RN125, RN122}. However, the previous polar codes-based IR schemes couldn't be accelerated by the efficient decoders without true random numbers (TRNs). In this article, we propose a novel polar codes-based IR scheme with frozen bits erasure (FBE) strategy.

Before IR procedure, Alice and Bob generate the sifted keys ($K_{s}^A$ and $K_{s}^B$) of length $n$ ($n=2^m$, $m\in N^+$) respectively, estimate the quantum bit error rate (QBER) as $E_\mu$ and pre-share the frozen vector $V$, which represents the positions of the $n-k$ frozen bits.
\newtheorem{myDef}{Definition}
\begin{myDef}
	$Ext(U,V)$ represents the vector which is composed of the element $u_i\in U$ when $v_i=1$, $v_i \in V$.
\end{myDef}

\subsection{Frozen Bits Erasure Strategy}

In our proposed frozen bits erasure (FBE) strategy, Alice calculates syndrome vector $W$ as 
\begin{equation}
	W=Encode(K_{s}^A) \wedge V = K_{s}^A G^n \wedge V\label{equ:W}.
\end{equation}
and sends the vector $W$ to Bob via the classical channel. $G^n$ is the generator matrix of polar codes and $G^nG^n$ equals to the identity matrix $I^n$.

Then, Alice generates a novel codeword $X$ as
\begin{equation}
	X=K_{s}^A \oplus Encode(W)\label{equ:X}.
\end{equation} 

At Bob's side, Bob receives the syndrome vector $W$. Then, Bob encodes the vector $W$ and generates the codeword with error bits as 
\begin{equation}
	X'=K_{s}^B \oplus Encode(W)\label{equ:X1}.
\end{equation}

According to Eq.~(\ref{equ:W}) and (\ref{equ:X}), the codeword $X$ can be further calculated as 
\begin{equation}
	\begin{aligned}
	X&=Encode(Encode(K_s^A) \wedge \overline{V})\\
		&=(K_s^A G^n \wedge \overline{V})G^n.\\
	\end{aligned}
\end{equation}

Encode $X$ to $U$, and $U$ equals to $K_s^A G^n \wedge \overline{V}$. Meanwhile, $X=Encode(U)$ for the $G^nG^n=I^n$. Therefore, the frozen bits of $U$ are 0, and the frozen bits in the decoding $X'$ procedure are 0. 

Finally, the vector $U=Encode(X)=K_s^A G^n \wedge \overline{V}$ equals to the decoded vector $U'$ from $X'$ when the decoding procedure is conducted successfully. And the users can also choose the information bits of $U$ and $U'$ as the output weak secure keys.
\begin{table}[H]
	\setlength\tabcolsep{3pt}
	\centering
	\caption{The features comparison of direct decoding (DD) strategy, bit flipping decoding (BFD) strategy and FBE strategy}\label{tab:comparison}
	\begin{tabular}{ccccc}
	\hline
	Strategy & Complexity & \makecell[c]{Syndrome\\bits} & \makecell[c]{Frozen\\bits} & \makecell[c]{With\\TRNs?}\\
	\hline
	DD & $O(n\log n)$ & $n-k$ & Variable & No\\
	BFD & $O(n\log n)$ & $n$ & Constant & Yes\\
	FBE & $O(n\log n)$ & $n-k$ & Constant & No\\
	\hline
	\end{tabular}
\end{table}
In the realistic implementation, Alice only needs to transmit $n-k$ frozen bits of $W$ via the classical channel and Bob can reconstructed $W$ from the received $n-k$ bits and the forzen vector $V$. Table~\ref{tab:comparison} shows the features comparison of DD strategy, BFD strategy and FBE strategy.

\subsection{Polar Codes-based IR Scheme with FBE strategy}

\begin{figure}[H]
	\centering
	\includegraphics[width=0.45\textwidth]{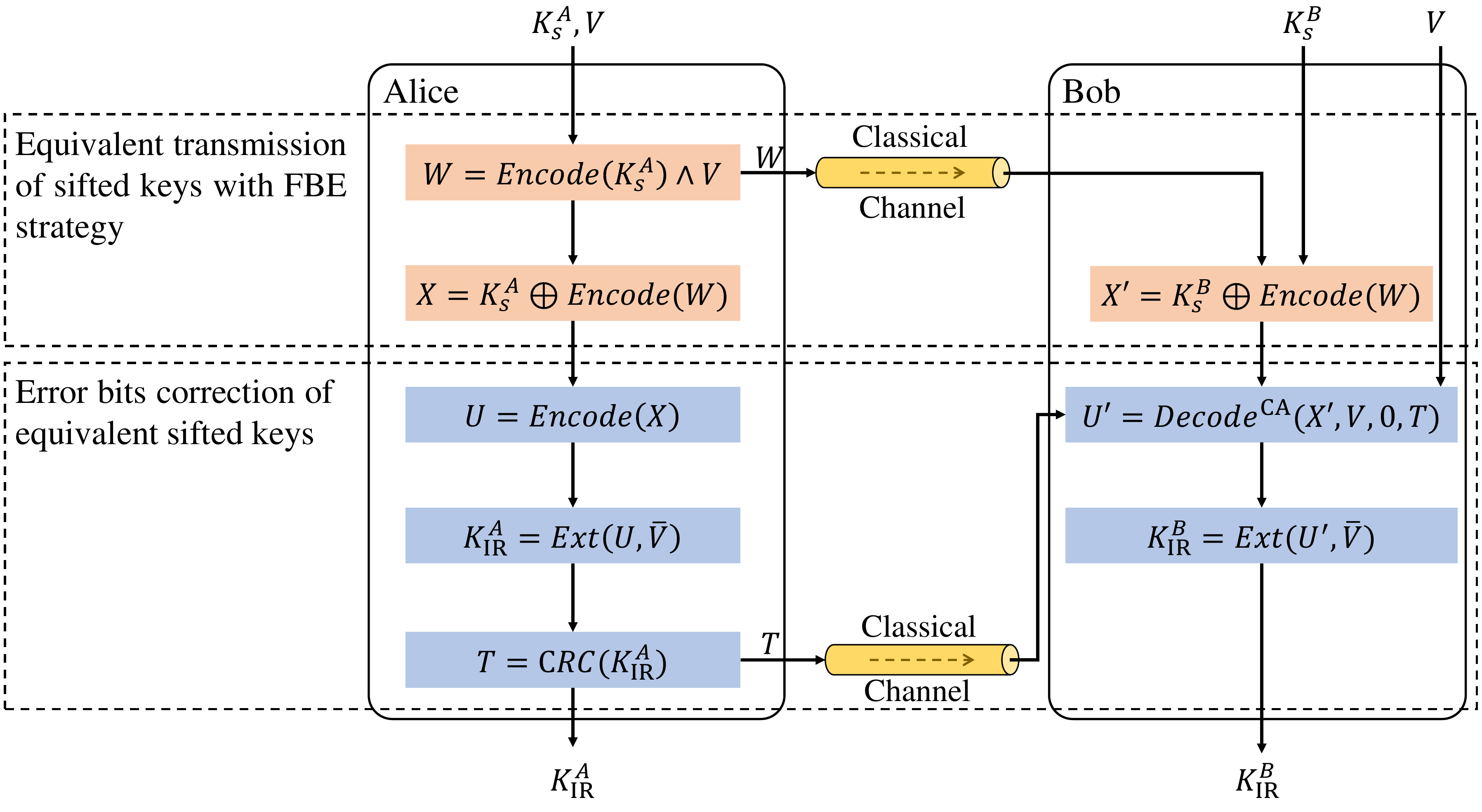}
	\caption{The diagram of the polar codes-based IR scheme with FBE strategy\label{fig:mainpro}. $Ext(U,\overline{V})$ reprents the vector composed of $\{u_i|v_i=1,u_i \in U, v_i \in V\}$. CRC($X$) represents the cyclic redundancy check value of vector $X$. $Decode^{\mathrm{CA}}(X,V,0,T)$ represents the decoded vector, which is generated from CRC-Aided (CA) decoding procedure when the codeword is $X$, the CRC value is $T$, the locations and values of frozen bits are $W$ and $U^W$, respectively.}
\end{figure}

Based on the proposed FBE strategy, we designed a novel polar codes-based IR scheme, which contains the two phases: equivalent transmission of sifted keys with FBE strategy and error bits correction of equivalent sifted keys. The diagram of this IR scheme is shown in Figure~\ref{fig:mainpro}. Before the IR procedure, Alice and Bob pre-share the CRC length $d$.

\subsubsection{Equivalent transmission of sifted keys with FBE strategy}
Alice distills the syndrome vector $W$ according to Eq.~(\ref{equ:W}) and calculates the codeword vector $X$ as Eq.~(\ref{equ:X}). Then, Alice sends the syndrome vector $W$ to Bob via classical channel. Bob receives the vector $W$ and calculates the codeword $X'$ as Eq.~(\ref{equ:X1}).

\subsubsection{Error bits correction of equivalent sifted keys}
Alice encodes codeword $X$ to vector $U$ and extracts the information bits of $X$ as $K_\mathrm{IR} ^A$. Then, Alice calculates the CRC value $T$ of $K_\mathrm{IR} ^A$ and sends $T$ to Bob through the classical channel. Bob performs a CRC-aided (CA) decoding procedure to decode $X'$ to $Z$ with the CRC value $T$ and frozen vector $V$. In the CA decoding procedure, the decoders are performed to generate $L$ temporary decoded vectors and the one that passed the CRC check is the final output decoded vector, same as the CRC-aided SCL decoder~\cite{RN188}. Finally, Bob extracts information bits of $U'$ as $K_\mathrm{IR}^B$.

In the realistic implementation, the procedures at Alice's side could be simplified as: (1) Encode $K_s^A$ to the vector $Y$. (2) Generate the syndrome vector $W=Y\wedge V$. (3) Extract the information bits of $Y$ as $K_\mathrm{IR} ^A=Ext(Y,\overline{V})$. (4) Calculate the CRC value $T=CRC(K_\mathrm{IR} ^A)$.

\section{PERFORMANCE ANALYSIS}
The polar codes-based IR scheme with FBE strategy has been implemented and a series of experiments have been conducted to evaluate its performance. In the experiments, the decoder is FSSCL decoder, the block size $n$ is \SI{1}{Mb}, the length of CRC is $32$ and the locations of frozen bits are determined by the optimized upgrading and degrading channels construction~\cite{RN106, myConstruction}. The sifted keys are collected from our reference frame independent QKD experiment and are extended to the targeted length and QBER~\cite{tang2021freerunning}. 

\begin{figure}[H]
	\centering
	\includegraphics[width=0.47\textwidth]{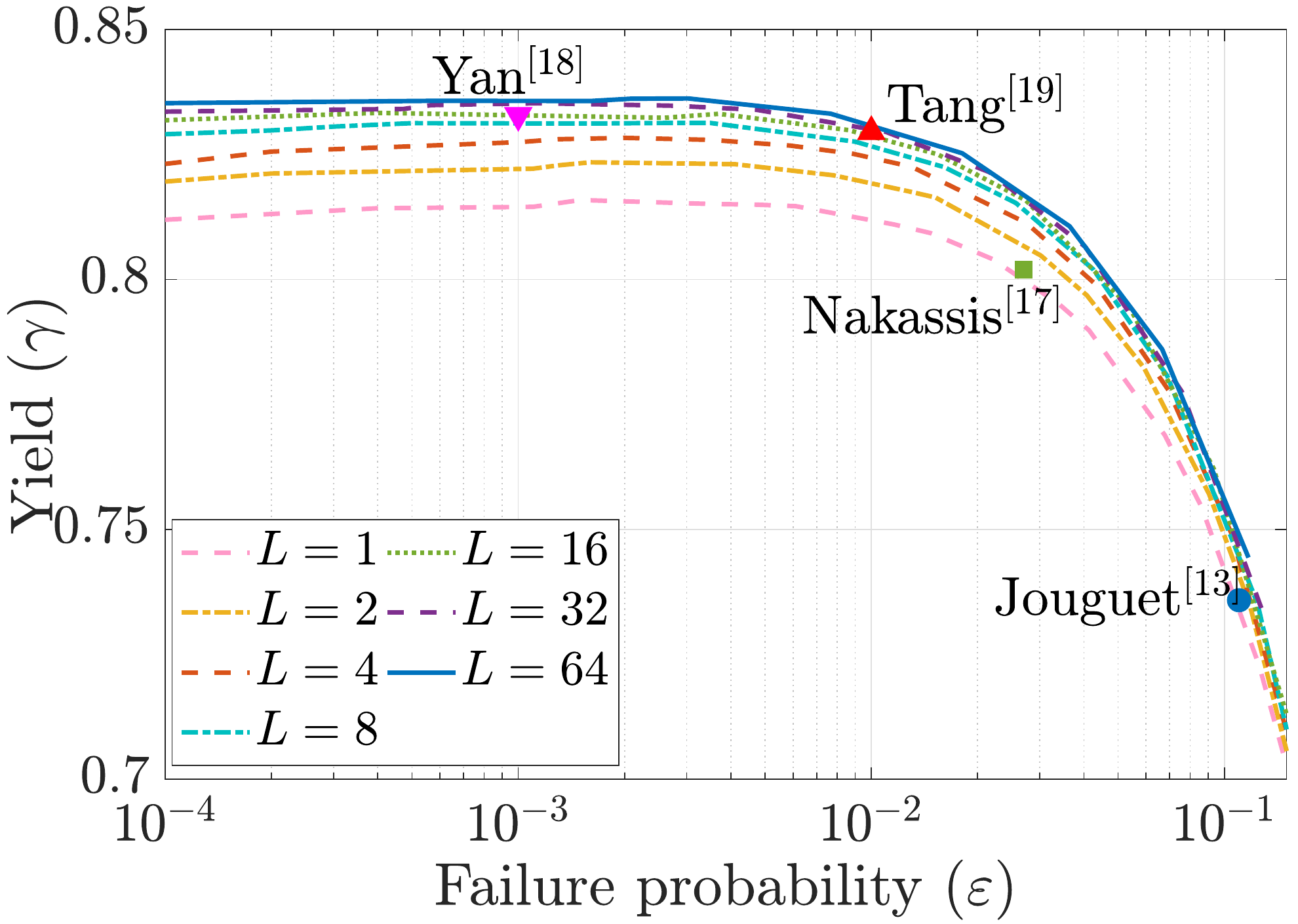}
	\caption{The yield of the polar codes-based IR schemes with FBE strategy against the failure probability, when decoder is FSSCL decoder, QBER equals 0.02, the list size $L$ is in $\{1,2,8,16,32,64\}$ and block size $n$ is \SI{1}{Mb}. The data of Tang's scheme is from the forward reconciliation phase in Ref.~\cite{RN676}.\label{fig:ferefficiency}}
\end{figure}

\subsection{Yield of IR}

The yield $\gamma$ of each sifted key bit represents the performance of IR schemes and is calculated from the failure probability $\varepsilon$ and the corresponding efficiency $f$ as Eq.~\ref{equ:gamma}. The lower bound of the efficiency $f$ is determined by the failure probability $\varepsilon$.

To evaluate the optimal yield of IR, we tested our polar codes-based IR scheme with FBE strategy for $10,000$ times each round with the list size $L \in \{1,2,8,16,32,64\}$, the QBER of $0.02$ and the failure probability from $10^{-4}$ to about $10^{-1}$. 

\begin{figure}[H]
	\centering
	\includegraphics[width=0.45\textwidth]{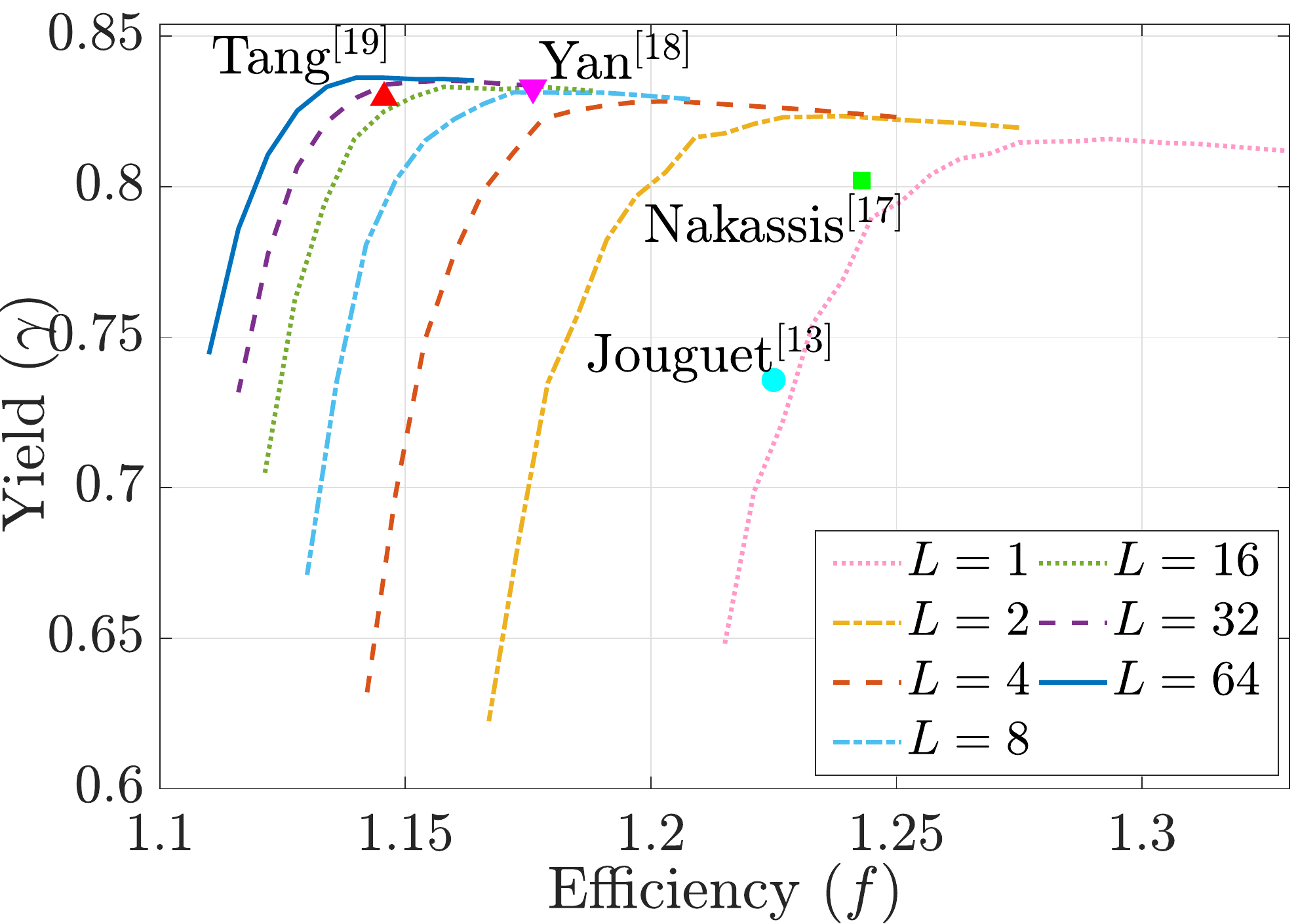}
	\caption{\label{fig:eff_gamma}The yield of the polar codes-based IR schemes with FBE strategy against the corresponding efficiency. The data of Tang's scheme is from the forward reconciliation phase in Ref.~\cite{RN676}.}
\end{figure}

Figure~\ref{fig:eff_gamma} shows the yield $\gamma$ against the failure probability $\varepsilon$. The increase of list size improves the yield of IR by improving the correction performance. The yield $\gamma$ first slowly increases to the maximum as the increase of failure probability $\varepsilon$ because the efficiency $f$ is reduced by the increase of $\varepsilon$. Then, the yield $\gamma$ decreases rapidly as the increase of failure probability $\varepsilon$ for the discard failure cases. Figure~\ref{fig:eff_gamma} shows the yield $\gamma$ against the efficiency $f$, which is corresponding to the failure probability $\varepsilon$.

The efficiency and the failure probability are shown in Table~\ref{L_and_f} of supplementary when the optimal yield is achieved and the list size $L \in \{1,2,8,16,32,64 \}$. Our scheme reaches the optimal yield of $0.8362$ with the list size of $64$, which is much higher than the conventional one-way polar codes-based IR schemes~\cite{RN99, RN98, RN129, RN676}. The performance of polar codes-based IR schemes could be further improved by appending the feedback procedure or improving the decoders, such as the scheme in Ref.~\cite{RN676}.

\subsection{Throughput of IR}

The ``bottleneck'' of IR is the decoding procedure at Bob's side. In this article, we use the throughput of the IR scheme at Bob's side to represent the whole IR scheme. 

\begin{figure}[H]
	\centering
	\includegraphics[width=0.45\textwidth]{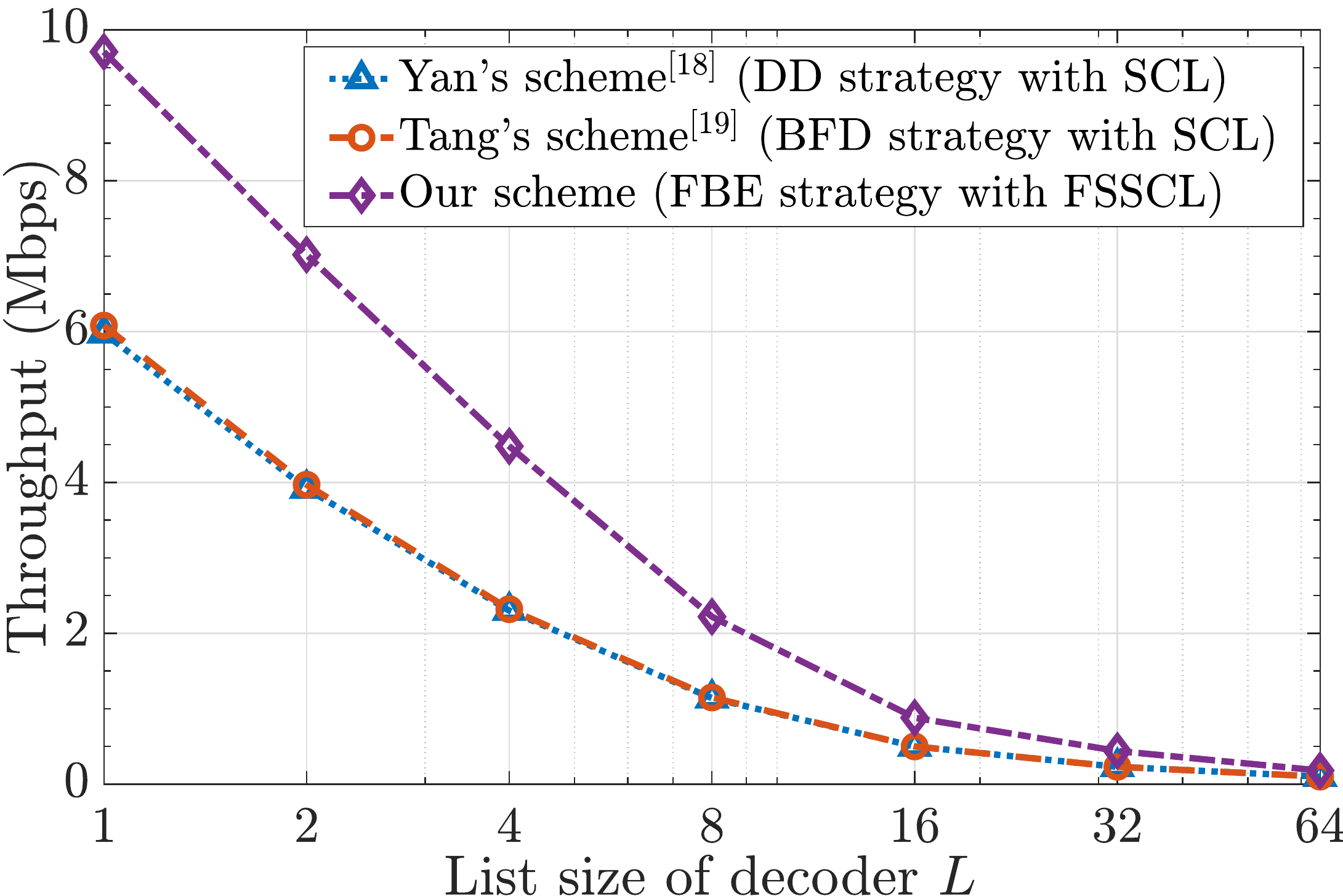}
	\caption{Throughput of the polar codes-based IR schemes with FBE strategy\label{fig:throughput}. DD strategy: direct decoding strategy, BFD strategy: bit flipping decoding strategy, FBE strategy: frozen bits erasure endcoding strategy.}
\end{figure}

We implemented the polar codes-based IR schemes in Ref.~\cite{RN676} (forward reconciliation) and Ref.~\cite{RN99}, which are based on the BFD strategy and DD strategy, respectively. The throughput of the implementations are evaluated with the block size of \SI{1}{Mb}, $E_\mu=0.02$, the list size $L\in \{ 1,2,4,8,16,32,64 \}$ and the efficiency in Table~\ref{L_and_f} of supplementary section. Figure~\ref{fig:throughput} shows the evaluated throughput result, and the detailed throughput of our scheme is shown in Table~\ref{L_and_f} of the supplementary section. The increase of list size improves efficiency but decreases the throughput. Our polar codes-based IR scheme reaches the throughput of 60\% higher throughput than the implementations of the previous schemes. Furthermore, the polar codes-based IR scheme with FBE strategy can apply the state-of-art decoder to reach the higher throughput and efficiency, such as the hardware-based decoders~\cite{RN107}.

Moreover, the proposed FBE strategy can be directly adapted into the existing polar codes-based IR schemes to adapt the improved SCL decoders without true random numbers~\cite{RN694, RN676}. And the proposed polar codes-based IR scheme is suitable in the practical QKD systems with the ultra worse link conditions, such as the satellite-to-ground QKD systems and drone-based QKD systems. 

\section{Conclusion}
In this article, we propose the polar codes-based information reconciliation (IR) scheme with the frozen bits erasure (FBE) strategy, where an equivalent transmission of the sifted keys is conducted, so that the frozen bits in the decoding procedure is erased to 0. Compared with the previous IR scheme, The polar codes-based IR scheme with the FBE strategy could be efficiently implemented on the commercial computer without the extra hardware, such as true random numbers generator. Furthermore, we implemented the scheme with the fast simplified successive cancellation list decoder. The implementation reaches the throughput of \SI{0.88}{Mbps} and the yield of $0.8333$ (efficiency of $1.760$ and failure probability of $0.0004$) with the decoder list size of $16$, the block size of \SI{1}{Mb} and QBER of $0.02$. Therefore, the IR scheme could be applied in the practical QKD systems, especially with the ultra worse link conditions.

\section*{Supplementary}

\subsection*{Polar Codes}
Polar codes, invented by Arikan in 2008, have the potential to reach the Shannon limit of binary discrete memoryless channel (B-DMC) in theory~\cite{RN122, RN125}. In polar codes, $n$ copies of B-DMC is polarized to a new set of bit-channels composed of $k$ error-free (``good'') channels and $n-k$ noisy (``bad'') channels. The positions of noisy channels are determined by the channel capacity, the Bhattacharyya parameter, or the error probability of each channel~\cite{RN125, RN106}. Frozen vector $V=[v_0,v_1,\cdots,v_{n-1}]$ is usually used to represent the positions of frozen bits, where $v_i=0$ ($v_i=1$) means the position $i$ is the information (frozen) bit.

In the encoding procedure of polar codes, $k$ infromation bits and $n-k$ pre-shared bits (usually $0$) are filled into error-free positions and noisy positions of $U$, respectively. Then, $U$ is encoded to the codeword $X$ as 
\begin{equation}
    X =UG^n= U  F^{\otimes \log n} B_n,
\end{equation}
where $F = \begin{bmatrix} 1 & 0\\1 & 1 \end{bmatrix}$, $B_n$ is the permutation matrix for bit-reversal operation and $G^nG^n$ is the identity matrix. Afterwards, the codeword $X$ is sent through the B-DMC. 

The receiver gets the measured codeword as $X '$ from B-DMC. The received codeword $X '$ could be decoded to $U$ with the pre-shared frozen bits. 

Arikan \textit{et al.} first proposed the successive cancellation (SC) decoder, whose complexity is $O(n \log n)$~\cite{RN125}. Then, the successive cancellation list (SCL) decoder is proposed to decrease the frame error rate with the complexity of $O(L n \log n)$, $L$ is the list size~\cite{RN392, RN188}. Especially, the efficient decoders, based on the pre-shared frozen bits, are developed from SC and SCL decoders with the same correction performance, \textit{e.g.} simplified successive-cancellation (SSC) decoder~\cite{RN534}, simplified successive cancellation list (SSCL) decoder~\cite{RN396}, fast simplified successive cancellation list (FSSCL) decoder~\cite{RN111}. The hardware-based decoders of polar codes with constant frozen bits are improved to reach the throughput of \SI{237}{Gbps}~\cite{RN107}.

\begin{figure}[H]
	\centering
	\includegraphics[width=0.47\textwidth]{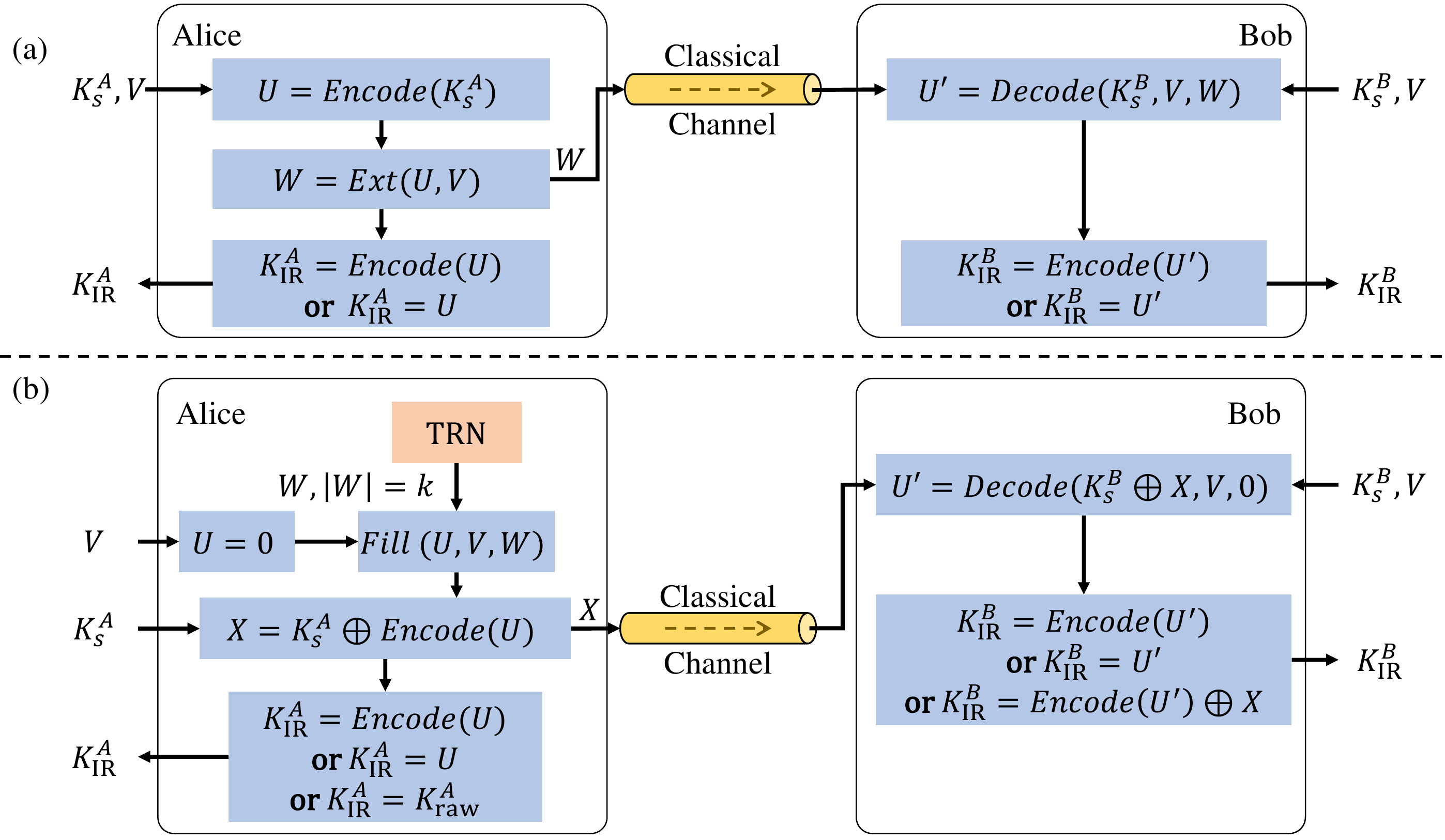}
	\caption{The diagram of polar codes-based IR schemes with DD strategy and BFD strategy: (a) DD strategy; (b) BFD strategy\label{fig:configuration}. $V$ is frozen vector pre-calculated from QBER $E_\mu$. TRN means true random number. $Encode(U)$ represents encoding $U$ to $UG^n$, $n$ is the length of $U$. $Ext(U,V)$ represents to extract the element $u_i$ when $v_i=1$, $u_i\in U$ and $v_i\in V$. $Fill(U,V,W)$ means filling $W$ into frozen bits of $U$ and $V$ is the frozen vector. $Decode(X,W,U^W)$ represents the decoded vector when the codeword is $X$, the frozen vector is $W$, the values of frozen bits are and $U^W$.}
  \end{figure}

\subsection*{Polar Codes-based Information Reconciliation Strategies}

Polar codes-based IR strategies guide how to adopt polar codes into IR schemes and mainly contain direct decoding (DD) strategy, bits flipping decoding (BFD) strategy, and length-adaptive BFD strategy~\cite{RN98}. The length-adaptive BFD strategy is suitable to any input length but has the lower efficiency than DD strategy and BFD strategy. In QKD systems, the input length of IR could be fixed to $2^m$ ($m\in N^+$), so that the polar codes-based IR schemes are mainly based on the DD strategy and the BFD strategy. Figure~\ref{fig:configuration} shows the diagram of DD strategy and BFD strategy.

In DD strategy, the frozen vector are pre-shared by the communication parties, and the length of sifted keys is $2^m$, $m\in N^+$. Alice encodes $K_{s}^A$ to $U$, sends the frozen bits of $U$ to Bob. Bob directly decodes $K_{s}^B$ to $U$ with the frozen bits of $U$ and encodes $U$ to $K_{s}^A$. Alice and Bob choose the information bits of $U$ or $K_{s}^A$ as $K_\mathrm{IR}$. The frozen bits are determined by $K_{s}^A$ and vary in each run. The DD strategy doesn't support the improved efficient decoders that require the constant frozen bits.

In BFD strategy, the input length is $2^m$, the frozen vector and the values of frozen bits are pre-shared by the communication parties. Alice generates a vector $U$ composed of the pre-shared frozen bits and the random bits. Afterwards, Alice calculates $X$ as $Encode(U) \oplus K_{s} ^A$ and sends $X$ to Bob. Bob decodes the vector $X\oplus K_{s} ^B$ to $U$ with the pre-shared frozen bits, encodes $U$ to $X$ and generates $K_{s}^A$. The BFD strategy has three choices of the weak secure key: $U$, $X$, and $K_{s}^A$. Meanwhile, the BFD strategy could be accelerated by the efficient decoders and is widely used in the recent polar codes-based IR schemes~\cite{RN676, RN694}. However, the generation of true random numbers (TRNs) would increase the complexity of systems and might open the security loopholes with inappropriate implementation. Meanwhile, $n$ syndrome bits are transmitted via public channel, which increases the overload of the public channel.

\begin{table}[H]
    \centering
    \caption{Experimental Environment Settings}\label{mytab}%
    \begin{tabular}{ccc}
        \hline
        &Parameters & Value\\
        \hline
        \multirow{4}{*}{\makecell[c]{Polar\\codes}}&Block size & \SI{1}{Mb}\\
        &CRC length & 32 \\
        &Decoder & FSSCL decoder~\cite{RN111}\\
        &Construction & \makecell[c]{Upgrading and\\degrading method~\cite{RN106}}\\
        \hline
        \multirow{6}{*}{\makecell[c]{Comp-\\uter}}&\makecell[c]{Operation Sys-\\tem} & Windows 10\\
        &CPU  & Intel I5-9300H\\
        &Cores per CPU & 4\\
        &Treads per core & 2\\
        &Memory & 16 GB\\
        &Compiler & Visual Studio 2019\\
        \hline
    \end{tabular}
\end{table}

\subsection*{Experimental settings}

Table~\ref{mytab} shows the experimental environment settings.

\begin{table}[H]
	\centering
	\caption{The optimal yield $\gamma$ and the throughput of our polar codes-based IR scheme with FBE strategy. $L\in \{1,2,4,8,16,32,64 \}$ and $E_\mu = 0.02$.\label{L_and_f}}     
	\begin{tabular}{ccccc}
	   \hline  
	  \makecell[c]{$L$} & \makecell[c]{$f$} & \makecell[c]{$\varepsilon$} &$\gamma$ &\makecell[c]{Throughput\\(Mbps)} \\ 
	  \hline
		1 & 1.293  &   0.0015 & 0.8159 & 9.71\\ 
		2 & 1.239 & 0.0016 & 0.8234 & 7.02\\ 
		4 &  1.202 & 0.0020 & 0.8283 & 4.48\\ 
		8 & 1.172 & 0.0035 & 0.8313& 2.22\\ 
		16 & 1.176 & 0.0004  & 0.8333& 0.88\\ 
		32 & 1.158  & 0.0011  &  0.8353& 0.44\\
		64 & 1.140 & 0.0030 & 0.8362&0.18 \\ 
		\hline   
	\end{tabular}
  \end{table}

\subsection*{Optimal yield}

Table~\ref{L_and_f} shows the The optimal yield and the throughput of our polar codes-based IR scheme with FBE strategy.

\section*{Acknowledgements}

This work was supported by National Natural Science Foundation of China under Grant No. 61972410, the Research Plan of National University of Defense Technology under Grant No. ZK19-13 and No. 19-QNCXJ-107 and the Postgraduate Scientific Research Innovation Project of Hunan Province under Grant No. CX20200003.

\section*{Additional Information}
The authors declare no conflicts of interest.

\bibliographystyle{unsrt}

\end{multicols} 
\end{document}